\documentclass[conference]{IEEEtran}

\usepackage{amsmath}
\usepackage{suffix}
\usepackage{mathtools}
\usepackage{cite}
\usepackage{color,soul}
 
\usepackage{multirow}
\usepackage{hhline}

\ifCLASSINFOpdf
\else
\fi

\usepackage{suffix,mathtools}
\DeclarePairedDelimiterX\MeijerM[3]{\lparen}{\rparen}%
{#3\,\delimsize\vert\begin{smallmatrix}#1 \\ #2\end{smallmatrix}}

\newcommand\MeijerG[8][]{%
  G^{\,#2,#3}_{#4,#5}\MeijerM[#1]{#6}{#7}{#8}}

\WithSuffix\newcommand\MeijerG*[7]{%
  G^{\,#1,#2}_{#3,#4}\MeijerM*{#5}{#6}{#7}}

\begin{document}

\title{Outage Probability Analysis of  RF/FSO-VLC Communication Relaying System}

\author{\IEEEauthorblockN{Milica Petkovic}
\IEEEauthorblockA{University of Novi Sad \\
Faculty of Technical Sciences \\
Novi Sad, Serbia\\
Email: milica.petkovic@uns.ac.rs}
\and
\IEEEauthorblockN{Aleksandra Cvetkovic}
\IEEEauthorblockA{University of Nis \\
Faculty of Electronic Engineering\\
Nis, Serbia\\
Email: aleksandra.cvetkovic@elfak.ni.ac.rs}
\and
\IEEEauthorblockN{Milan Narandzic}\IEEEauthorblockA{University of Novi Sad \\
Faculty of Technical Sciences \\
Novi Sad, Serbia\\
Email: orange@uns.ac.rs}}

\maketitle

\begin{abstract}
This paper presents an analysis of the asymmetric relaying system which provides communication between hybrid outdoor sub-system and indoor visible light communications (VLC) access points. The outdoor sub-system represents a hybrid radio-frequency (RF)/free-space optical (FSO) system, introduced to reduce the impact of weather conditions on transmission quality. Closed-form outage probability analytical expression is derived. Numerical results are presented  and confirmed by Monte Carlo simulations. The effects of system and channel parameters on the outage probability performance are investigated and discussed. Greater optical transmitted power of the VLC sub-system reflects in better system performance, as well as lower indoor environment height. When the indoor room is higher, the propagation of the optical signal is longer, and there will be greater power dissipation and performance deterioration. In addition, the outage probability floor is noticed, which is important limiting factor for relaying system design.
\end{abstract}


\IEEEpeerreviewmaketitle

\section{Introduction}
The optical wireless communication (OWC) is a modern, attractive  communication technique, which receives an attention due to benefits of both wireless and optical technologies. The OWC systems are   licence-free and low cost, offering high data-rates and very secured transmission. The OWC finds its applications in many areas, such as short-distance interconnects on integrated circuits, indoor communications, terrestrial outdoor links, satellite links, etc. \cite{survey,OWC_MATLAB,book2}.

The indoor OWC systems, also known as visible light communications (VLC), represent the transmission of the optical signal from visible spectrum. The VLC systems  often employ light emitting  diodes (LEDs),  for both illumination and data communications \cite{OWC_MATLAB, survey,bookvlc,vlc1}. As the OWC application refers to the outdoor enviroments, the free-space optics (FSO) represents terrestrial point-to-point link, which usually employs laser optical source to perform transmission in infra-red  band. The FSO systems represent highly-secured and low-cost wireless transmission. But, the application of the FSO systems is limited by line-of-sight (LOS) requirement, as well as weather and turbulence conditions of the atmospheric channel \cite{survey,OWC_MATLAB,book2}.

Recently,  Gupta \textit{et al.} \cite{model}  presented a statistical analysis of the cascaded relaying FSO-VLC system, where the information is delivered to the end users via indoor multiple VLC access points. Motivated by this study, we expand the analysis with the fact that the outdoor FSO link is highly sensitive on conditions  of the atmospheric medium. It is known that the quality of the FSO signal transmission is highly dependent on the visibility, thus it is adversely impacted by fog. On the other hand, the RF transmission is little impacted by fog and mostly affected by rain  \cite{survey,OWC_MATLAB}. For that reason, we consider the first sub-system to be hybrid RF/FSO link \cite{hyb1,hyb2}, employing selection combining (SC) diversity technique to select active link. Hence, the realization of the outdoor signal transmission is dependent on the RF and FSO channel states, which show different sensitivity to snow and rain. 
The indoor traffic is managed by multiple VLC
access points, which distribute data to the end mobile users. Each end user is connected with a single VLC access point, which provides the best channel gain for considered user.
The decode-and-forward (DF) relay is assumed to integrate hybrid RF/FSO link and  indoor VLC access points.
For considered system, the outage probability  analytical expression is derived, in order to perform analysis of the system and channel parameters on the transmission quality.
Derived  expression is utilized to obtained numerical results, which are further validated by Monte Carlo simulations.

The rest of the paper is organized as follows. Section II presents the system and channel model. The outage probability analysis is given in section III, while numerical results are shown in section IV. Some concluding remarks are listed  in section V. 
 
\section{System and channel model}

\begin{figure}[!b]
\centering
\includegraphics[width=3.2in]{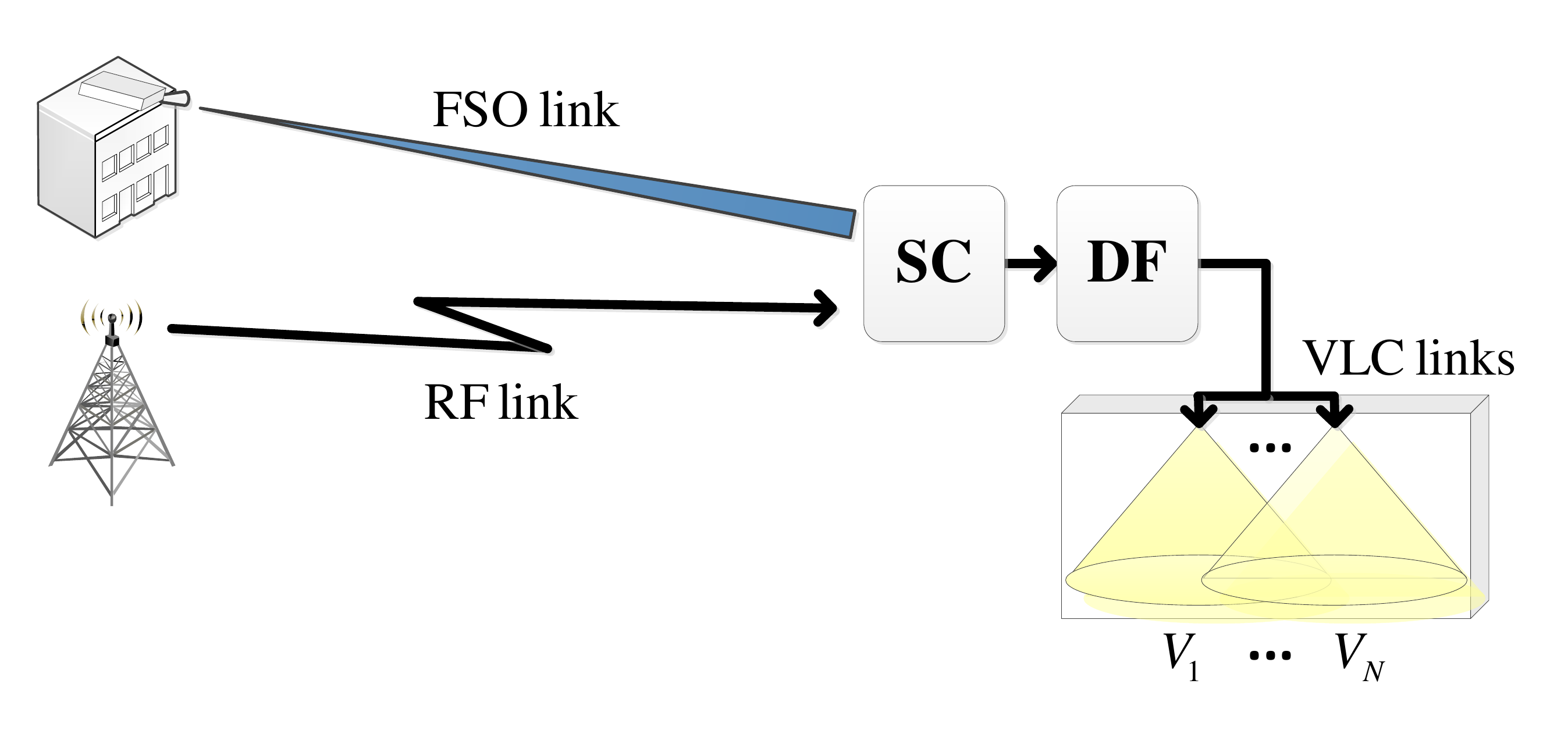}
\caption{System model of a RF/FSO-VLC  system: SC selects between RF and FSO, DF relay decodes data and forwards it to all VLC LED sources.}
\label{Fig1}
\end{figure}

As it is shown in Fig.~\ref{Fig1}, the first part of asymmetric relaying system represents the hybrid RF/FSO sub-system, with SC spatial diversity technique that processes only the one of the link outputs. The active link will be the one with better channel conditions (greater signal-to-noise ratio (SNR)), i.e., the selection of the active link is dependent on weather conditions. 
The DF relay performs retransmission on the visible light frequency via  $ N $   VLC access points placed in some indoor environment. The multiple LEDs, denoted by $ V_n $, $ n=1,\ldots,N $, are positioned on the ceiling of the room to  transmit the information data to the end users, which are  uniformly distributed over the coverage area of the room.  

The fading over the RF channel experiences Rayleigh distribution. Thus the instantaneous SNR, $\gamma _{RF}$, is exponentially distributed random variable with the  probability density function (PDF)  given by  \cite{Simon}
\begin{equation}
f_{\gamma _{RF}}\left( \gamma  \right) = \frac{1}{\mu_{1}}e^{ - \frac{\gamma }{\mu_{1}}},
\label{pdfRF}
\end{equation}
where $\mu_1 $ is the average SNR over RF  link.
The cumulative density function (CDF) of the instantaneous SNR,  $\gamma _{RF}$, is given by \cite{Simon}
\begin{equation}
F_{\gamma_{RF}}\left( \gamma  \right) = 1 - e^{ - \frac{\gamma }{\mu _{1}}}.
\label{cdfRF}
\end{equation}

Furthermore, it is assumed that the intensity fluctuations of the received optical
signal are modeled by Gamma-Gamma distribution, convenient in wide range of atmospheric turbulence conditions. Thus, the PDF of the  instantaneous SNR, $\gamma _{FSO}$, is defined as \cite{hyb1}
\begin{equation}
f_{\gamma_{FSO}}\left( \gamma  \right) = \frac{{\left( \alpha \beta \right)}^{\frac{\alpha  + \beta }{2}}{\gamma ^{\frac{\alpha  + \beta }{4} - 1}}}
{{\Gamma \left( \alpha  \right)\Gamma \left( \beta  \right)\mu_2^{\frac{{\alpha  + \beta }}{4}}}}
K_{\alpha  - \beta}\left( {2\sqrt {\alpha \beta \sqrt {\frac{\gamma }{\mu _2}} } }\right),
\label{pdfFSO}
\end{equation}
where $ \Gamma \left( \cdot  \right) $ is the gamma function defined by \cite[(8.310.1)]{grad},  $K_\nu \left( \cdot  \right) $ is the $\nu $-th order modified Bessel function of the second kind defined by \cite[(8.432.2)]{grad}, and $\mu_2 $ represents the electrical SNR over FSO link.
With assumption of the Gaussian plane wave propagation and zero inner scale, the parameters $ \alpha $ and $  \beta$ are defined as $ \alpha~=~( \exp {[ 0.49 \sigma _R^2 ( 1 + 1.11 \sigma _R^{12/5} )^{-7/6} ]} - 1 )^{-1}$ and $ \beta~=~( \exp {[ 0.51 \sigma _R^2 ( 1 + 0.69 \sigma _R^{12/5} )^{-5/6}]} - 1)^{-1} $,
where the Rytov variance, denoted by  $ \sigma _R^2$, is used as a metric for atmospheric turbulence strength \cite{book2,OWC_MATLAB}.

The corresponding CDF of  $\gamma _{FSO}$ is given by \cite{hyb1}
\begin{equation}
F_{\gamma _{FSO}}\left( \gamma  \right) = \frac{1}{\Gamma \left( \alpha  \right)\Gamma \left( \beta  \right)}
G_{1,3}^{2,1}\left( {\alpha \beta \sqrt {\frac{\gamma }{\mu _2}} \left| \!\!\!\!\!\! {\;\begin{array}{*{20}{c}}
1\\
{\begin{array}{*{20}{c}}
{\alpha,}\!&\!{\beta ,}\!&\!0
\end{array}}
\end{array}} \right.} \!\!\!\!\!\!\!\right),
\label{cdfFSO}
\end{equation}
where $G_{p,q}^{m,n}\left( \cdot  \right) $ is the Meijer's \textit{G}-function \cite[(9.301)]{grad}.

The instantaneous SNR of the hybrid RF/FSO sub-system, i.e., SNR at the SC output, is defined  as
\begin{equation}
\gamma_1 = \max \left( \gamma _{RF}, \gamma _{FSO}  \right),
\label{snr1}
\end{equation}
while the CDF of $\gamma_1$ is determined as
\begin{equation}
F_{\gamma_1}\left( \gamma  \right) = F_{\gamma _{RF}}\left( \gamma  \right)F_{\gamma _{FSO}}\left( \gamma  \right),
\label{cdf1}
\end{equation}
where the CDFs $F_{\gamma _{RF}}\left( \gamma  \right)$ and $ F_{\gamma _{FSO}}\left( \gamma  \right)$ are previously defined in (\ref{cdfRF}) and (\ref{cdfFSO}), respectively.

Regarding the  indoor VLC  sub-system, it is assumed that the energy of  LOS component is greater than the energy of all reflected signals \cite{OWC_MATLAB, vlc1}. The reflected signals can be ignored,
thus every end user accepts only the strongest  signal  sent from  all $N$ LEDs. 
Further, the optical signal transmission to end user is performed by one of  $N$ VLC links, which  has the best channel conditions. The intercell interferences received  from other LEDs are  approximated with additive Gaussian process \cite{model, vlc2}.

\begin{figure}[!b]
\centering
\includegraphics[width=2.2in]{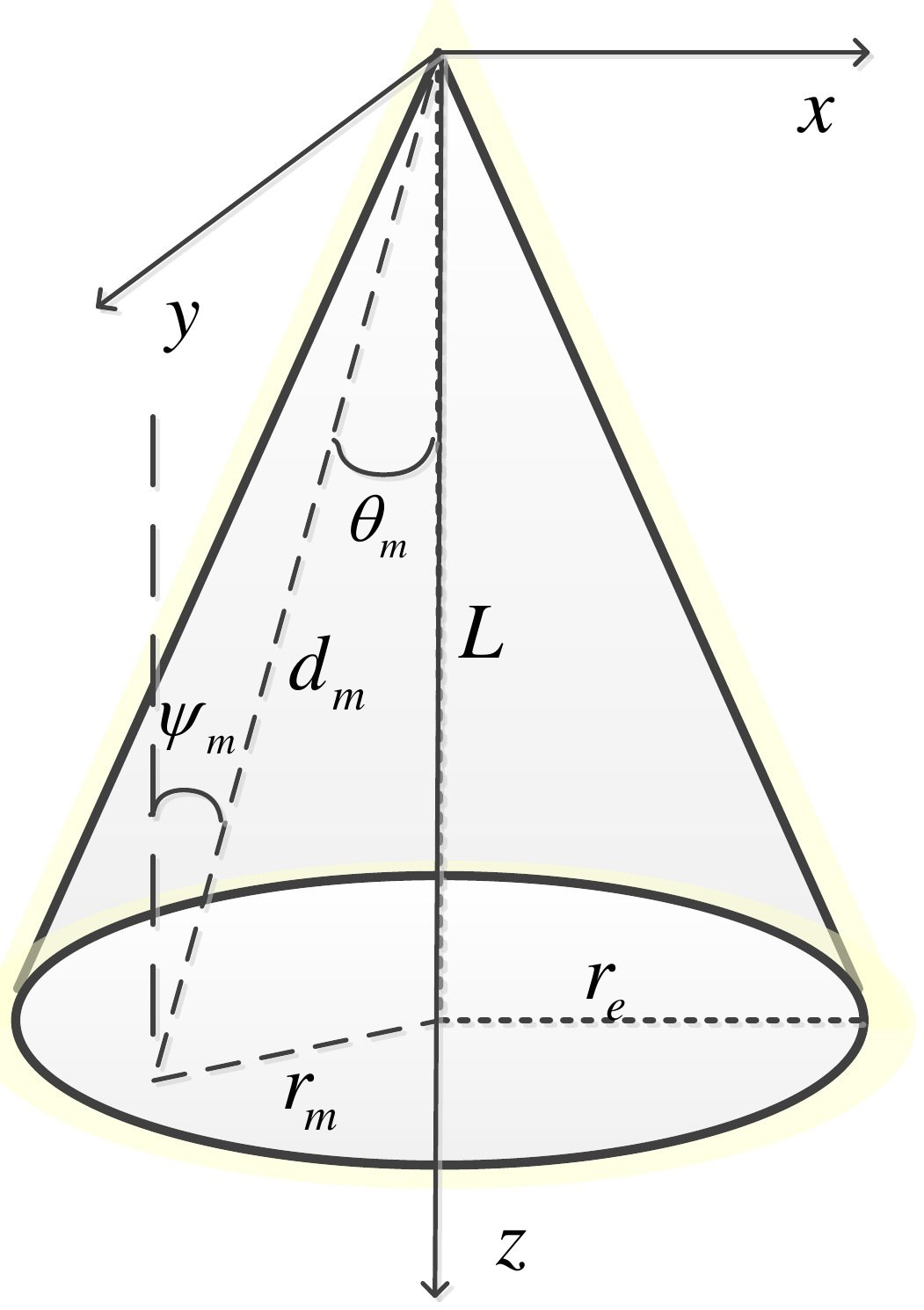}
\caption{Geometry of the LOS VLC propagation model}
\label{Fig2}
\end{figure}

The  LED transmitter is placed at height $ L $ from the $ m $-th end user, with the angle of irradiance $\theta_m$,   the angle of incidence $\psi _m$, and radius $r_m$  on the polar coordinate plane, as it can be observed in Fig.~\ref{Fig2}. The Euclidean distance between the  VLC access point and the $ m $-th photodetecor receiver is denoted by $ d_m $.
Furthermore, the semi-angle at the half-illuminance of LED,    $\Phi_{1/2}$, is in relation with  the maximum radius of a LED cell footprint,   $r_e$, as $r_e = L\sin \left( \Phi_{1/2}  \right)/\cos \left( \Phi_{1/2} \right)$.
The order of Lambertian emission is defined as 
\cite{vlc1, OWC_MATLAB,vlc3}
\begin{equation}
m_l =  - \frac{\ln 2}{\ln \left( \cos \Phi_{1/2} \right)}.
\label{m}
\end{equation}

The channel gain of the  LOS link between LED and the $ m $-th end user terminal is given by \cite{vlc1, vlc3}
\begin{equation}
I_m = \frac{A\left( m_l + 1\right)\Re}{2\pi d_m^2}\cos ^{m_l}\left( \theta _m \right)Tg\left( \psi _m \right)\cos \left( \psi _m \right), 
\label{I_n1}
\end{equation}
where $ A $ is a photodetector physical surface area, $\Re$ represents  a photodetector responsivity, $T$ is  the gain of the optical filter, and $g\left( \psi_m \right)$ the optical concentrator defined as  \cite{vlc2}
\begin{equation}
g\left( \psi _m \right) = \left\{ \begin{array}{*{20}{c}}
{n^2/  \sin^2\left( \Psi \right),\quad 0\leq\psi _m \leq \Psi}\\
{0,\quad \psi _m > \Psi}
\end{array} \right.  ,
\label{g_n}
\end{equation}
where $n$ is the refractive index of lens at a photodetector and  the field of view (FOV) of the photodetector receiver is  $\Psi$.
With assumption that the surface of  photodetector receiver is parallel to the ground, as well as that photodetector receiver has no orientation towards the LED, it holds that $\theta_m\!=\! \psi_m$.
Based on  relations $d_m =\! \sqrt {r_m^2 + L^2} $,  and  $\cos \left( \theta_m \right)\!=\!\frac{L}{\sqrt {r_m^2 + L^2} }$,  $\Im   = \frac{A\left( m_l + 1 \right)\Re}{2\pi}Tg\left( \psi _m \right)L^{m_l + 1}$ observed in Fig.~\ref{Fig2}, the channel gain is rewritten as
\begin{equation}
I_m = \frac{\Im }{\left( r_m^2 + L^2 \right)^{\frac{m_l + 3}{2}}}.
\label{I_n2}
\end{equation}
Under the assumption that user position is random over circular area,  the radial distance is modeled by a uniform distribution with the PDF 
\begin{equation}
f_{r_m}\left( r \right) = \frac{2r}{r_e^2},
\label{pdf_rn}
\end{equation}
defined in $0 \le r \le r_e $, while  the PDF of the channel gain $ I_m $ is derived as
\begin{equation}
f_{I_m}\left( I \right) = \frac{2}{r_e^2\left( m_l + 3 \right)} \Im ^{\frac{2}{m_l + 3}}I^{ - \frac{m_l + 5}{m_l + 3}},
\label{pdf_In}
\end{equation}
defined in $ I \in \left[ I_{\min },I_{\max } \right] $, where $I_{\min} = \frac{\Im}{{\left( r_e^2 + L^2 \right)}^{\frac{m_l + 3}{2}}}$ and  $I_{\max} = \frac{\Im}{L^{m_l + 3}}$.
The instantaneous SNR of the VLC channel of the $ m $-th end user can be defined as
\begin{equation}
\gamma_{VLC_m} = \frac{P_t^2 \eta ^2I_m^2}{N_0B},
\label{snrVLC}
\end{equation}
where  $P_t$ is transmitted optical  power of each LED, $\eta$ is an optical-to-electrical  conversion efficiency, $N_0$ denotes  noise spectral density, and $B$ the baseband modulation bandwidth.

Based on (\ref{pdf_In}) and (\ref{snrVLC}), considering  $\mu_{VLC} = \frac{P_t^2\eta ^2}{N_0B}$, the PDF of the instantaneous SNR of the $ m $-th end user  is  derived as
\begin{equation}
f_{\gamma_{VLC_m}}\left( \gamma  \right) = \frac{\mu_{VLC}^{\frac{1}{m_l + 3}} \Im^{\frac{2}{m_l + 3}}}{r_e^2\left( m_l + 3 \right)}\gamma^{ - \frac{m_l + 4}{m_l + 3}},
\label{pdfn}
\end{equation}
defined in $ \gamma \in \left[ \gamma_{\min },\gamma_{\max } \right] $, where $\gamma_{\min} = \frac{\mu_{VLC}  \Im^2}{{\left( r_e^2 + L^2 \right)}^{\frac{m_l + 3}{2}}}$ and  $\gamma_{\max} = \frac{\mu_{VLC} \Im^2}{L^{2 \left(m_l + 3\right)}}$. 
The corresponding CDF of the instantaneous SNR of the $ m $-th end user  is derived as
\begin{equation}
F_{\gamma_{VLC_m}} \left( \gamma  \right) = 1 + \frac{L^2}{r_e^2} - \frac{\Im^{\frac{2}{m_l + 3}}}{r_e^2}\left( \frac{\gamma }{\mu_{VLC}} \right)^{ - \frac{1}{m_l + 3}}.
\label{cdfn}
\end{equation}
The VLC link that performs the optical signal transmission will be established between the mobile user and the closest LED.
With assumption that  all VLC links are independent and identically distributed random variables, the CDF of the best instantaneous SNR regarding the $ m $-th end user, denoted by $\gamma_2$, is 
\begin{equation}
F_{\gamma _2}\left( \gamma  \right) = [F_{\gamma_{VLC_m}}\left( \gamma  \right)]^N,
\label{cdf2}
\end{equation}
where  $F_{\gamma_{VLC_m}}\left( \gamma  \right)$ is the CDF of the instantaneous SNR of the VLC link between the $ n $-th LED and the $ m $-th end user defined in (\ref{cdfn}). 

\section{Outage probability analysis}

For asymmetric DF based RF/FSO-VLC system, the equivalent end-to-end SNR at the $ m $-th end user is defined as
\begin{equation}
\gamma_{eq} = \min \left( \gamma _1,\gamma_2 \right),
\label{SNReq}
\end{equation}
where $\gamma _1$  is the instantaneous SNR of the hybrid RF/FSO sub-system and $\gamma _2$   is the instantaneous SNR of the active VLC link.
The outage probability for a single end user is determined  as the probability that $ \gamma_{eq} $ falls below a predetermined outage protection value, $ \gamma_{th} $, which can be written as
\begin{equation}
\begin{split}
P_{out}& = \Pr \left[ \gamma_{eq} < \gamma _{th} \right] = \Pr \left[ \min \left( \gamma _1,\gamma _2 \right) < \gamma _{th} \right] \\
& = F_{\gamma _1}\left( \gamma _{th} \right) + F_{\gamma _2}\left( \gamma _{th} \right) - F_{\gamma _1}\left( \gamma _{th} \right)F_{\gamma _2}\left( \gamma _{th} \right)
\end{split},
\label{Pout}
\end{equation}
where the CDFs   $F_{\gamma _1}\left(  \cdot  \right)$ and  $F_{\gamma _2}\left(  \cdot  \right)$  are previously defined in (\ref{cdf1})  and (\ref{cdf2}), respectively.

\section{Numerical results and discussions}

\begin{table}[b]
\centering
\caption{\bf Constants and system parameters}
\begin{tabular}{ccc}
\hline
name  & symbol & value= \\
\hline
photodetector surface area & $ A $  & $ 1~{\rm cm}^2 $\\
responsivity         & $\Re$  & $0.4~{\rm A}/{\rm W}$\\
optical filter  gain & $T$ & $1$ \\
refractive index of lens at a photodetector & $n$ & $1.5$ \\
FOV of receiver & $\Psi$ & $60^{\circ}$ \\
optical-to-electrical conversion efficiency  & $\eta$ & $0.8$ \\
noise spectral density & $N_0$ & $10^{-21}~{\rm W}/{\rm Hz}$ \\
Rytov variance for weak \\
 atmospheric turbulence condition & $\sigma_R^2$ & $0.25~(<1)$ \\
Rytov variance for strong \\
 atmospheric turbulence condition & $\sigma_R^2$ & $2~(>1)$ \\
\hline
\end{tabular}
  \label{table}
\end{table} 

In this section, numerical results are presented  and validated by Monte Carlo simulations. The values of the system and channel parameters are given in Table I \cite{vlc3,vlc4}. Numerical results are obtained based on the outage probability expression in (\ref{Pout}), which includes the CDF expressions presented in (\ref{cdf1})  and (\ref{cdf2}).

\begin{figure}[!t]
\centering
\includegraphics[width=3.5in]{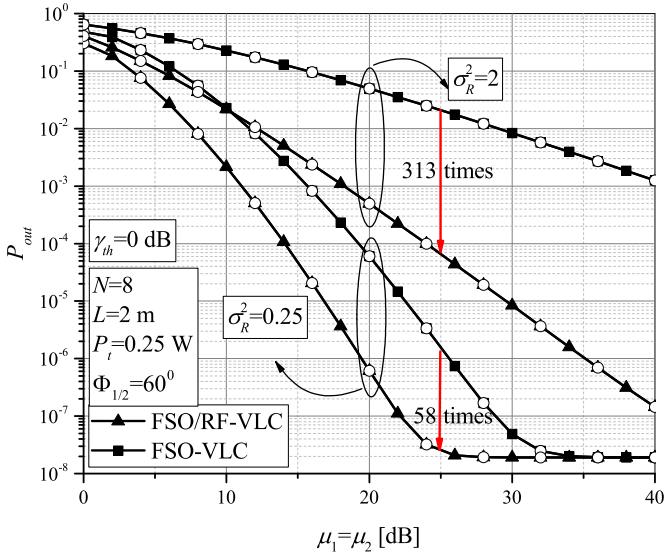}
\caption{Outage probability for RF/FSO-VLC and FSO-VLC systems in various atmospheric turbulence strengths}
\label{Fig3}
\end{figure}

The outage probability in the function of $\mu_1=\mu_2$ is presented in Fig.~\ref{Fig3} for two system scenarios. Besides 
considered system where the first part is hybrid RF/FSO system, the results for scenario analyzed in \cite{model} are also presented, referring to the case where the first sub-system includes only a single FSO link. 
Weak ($\sigma_R^2 = 0.25$) or  strong ($\sigma_R^2 = 2$) atmospheric turbulence conditions are considered. From presented results, it is clear that the system performance can be significantly improved by introducing the RF link as a backup link for the FSO system. This improvement is highly expressed when the FSO link suffers from strong atmospheric turbulence. For example, if $\mu_1=\mu_2=25~{\rm dB}$, by adding the RF link besides the FSO one with SC technique, the outage probability is improved about 58  times when $\sigma_R^2 = 0.25$, while about 313 times when $\sigma_R^2 = 2$.

\begin{figure}[!t]
\centering
\includegraphics[width=3.5in]{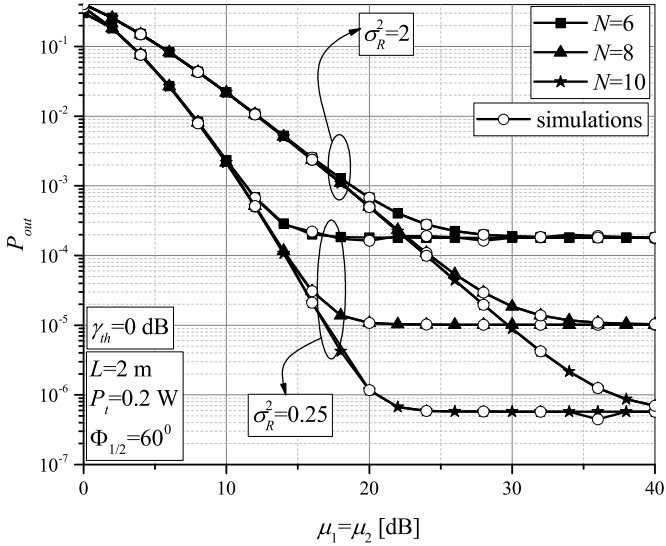}
\caption{Outage probability vs. $\mu_1=\mu_2$ for different number of the VLC access points and various atmospheric turbulence strengths}
\label{Fig4}
\end{figure}

Fig.~\ref{Fig4}  presents the outage probability dependence on $\mu_1=~\mu_2$ when the FSO link is affected by different atmospheric turbulence conditions. The number of the VLC access points can be 6, 8 or 10. As it is expected, system performs better when the FSO link suffers from weak  turbulence conditions. Furthermore, in the range of medium and high $\mu_1=\mu_2$ (greater RF signal power and greater FSO optical signal power), the overall system performance improvement is noticed when the  number of available VLC channels is greater. In the range of low values of $\mu_1=\mu_2$, the number of the VLC channels will not have impact on the system performance. In this scenario, the first sub-system will be in outage due to low transmitted power, thus the second system conditions improvement will not affect overall system performance.

\begin{figure}[!t]
\centering
\includegraphics[width=3.5in]{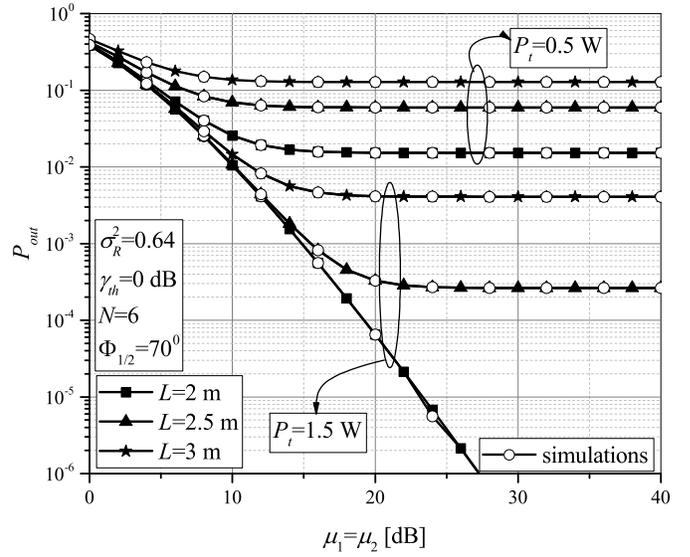}
\caption{Outage probability vs. $\mu_1=\mu_2$ for different transmitted optical power and different room heights}
\label{Fig5}
\end{figure}

Outage probability dependence on $\mu_1=\mu_2$ for different values of LED output optical power, as well as different indoor environment height, is depicted in Fig.~\ref{Fig5}. With greater LED radiation of the optical power, system performs better, but the cost and power consumption of the system is greater. It can be observed that the system performance is improved with lower dimension of the room height. When $L$ is greater, the propagation path of the optical signal is longer, thus the total received power at the photodetector is reduced.

In both Fig.~\ref{Fig4}  and Fig.~\ref{Fig5} the existence of the outage probability  floor is noticed in the range of greater values of  $\mu_1=\mu_2$. When  transmitted LED output power is constant, at some point, further increase of the RF or FSO signal power will not lead to the  outage  performance improvement. This outage probability floor appears at lower values of  $\mu_1=\mu_2$ when LED output power is lower, when the indoor space height is greater, as well as when $N$ is lower and in weak  turbulence conditions. This outage floor is limiting factor, thus, having high importance for RF/FSO-VLC system design.

\section{Conclusion}
In this paper, the analysis of the asymmetric relaying system, which was used to ensure traffic delivery to end users  by indoor VLC multiple access points, was  presented. The  outdoor sub-system was  hybrid RF/FSO system with SC diversity technique. 
 Closed-form outage probability was derived and used to obtained numerical results, which were validated by Monte Carlo simulations. Various effects of system parameters on the outage probability performance were discussed. 
 
It was presented that considered system performs better than the system when the first sub-system is a single FSO link, especially in strong atmospheric turbulence conditions. Number of available VLC access points determine the system performance to large extent. The outage probability floor was noticed, which is of high importance for system design and implementation. It was proved that greater LED output power improves system performance,  but in expense of the energy and cost savings.

\section*{Acknowledgment}
This work has received funding from the European Union Horizon 2020 research and innovation programme under the Marie Skodowska-Curie grant agreement No 734331.
The paper was supported in part by the Ministry of Education,
Science and Technology Development of the Republic of
Serbia under grant TR-32035.


\begin{thebibliography}{1}
\IEEEtriggeratref{5}

\bibitem{book2}
L. C. Andrews and R. N. Philips, \emph{Laser Beam Propagation through Random Media}. 2nd ed., Washington, USA: Spie Press, 2005.

\bibitem{OWC_MATLAB}
Z. Ghassemlooy, W. Popoola, and S. Rajbhandari, \textit{Optical Wireless Communications: System and Channel Modelling With MATLAB$ ^ {\textregistered}$}. Boca Raton, FL, USA: CRC Press, 2013.

\bibitem{survey}
M. A. Khalighi and M. Uysal, "Survey on free space optical communication: A communication theory perspective," \emph{IEEE Commun. Surveys Tuts.,} vol. 16, no. 4, pp. 2231--2258, Fourthquarter 2014.

\bibitem{bookvlc}
Z. Ghassemlooy, L. N. Alves, S.  Zvanovec, and M. A. Khalighi, (Eds.). \emph{Visible Light Communications: Theory and Applications}.  Boca Raton, FL, USA: CRC Press, 2017.

\bibitem{vlc1}
T.~Komine and M.~Nakagawa, "Fundamental analysis for visible-light
communication system using LED lights," \emph{IEEE Trans.
Consum. Electron.,} vol. 50, no. 1, pp. 100--107, Feb. 2004.


\bibitem{model}
A. Gupta, N. Sharma, P. Garg, and M.-S. Alouini,  "Cascaded FSO-VLC communication system," \emph{IEEE Wireless Commun. Lett.},  vol. 6, no. 6,  pp. 810--813, Dec. 2017.


\bibitem{hyb1}
N. D. Chatzidiamantis, G. K. Karagiannidis, E. E. Kriezis, and
M. Matthaiou, "Diversity combining in hybrid RF/FSO systems
with PSK modulation," in \textit{Proc. 2011 ICC}, Kyoto, pp. 1--6.

\bibitem{hyb2} 
B. He and R. Schober, "Bit-interleaved coded modulation for
hybrid RF/FSO systems," \emph{IEEE Trans. Commun.}, vol. 57, no.
12, pp. 3753--3763, 2009.


\bibitem{Simon}
M. K. Simon and M.$ - $S. Alouni, \textit{Digital Communication over Fading Channels}. 2nd ed., New York, NY: John Wiley \& Sons Inc., 2004.

\bibitem{grad}
I. S. Gradshteyn and I. M. Ryzhik, \textit{Table of Integrals, Series, and Products}. 6th ed., New York: Academic, 2000.

\bibitem{vlc2}
D. A. Basnayaka and H. Haas, "Design and analysis of a hybrid
radio frequency and visible light communication system," \emph{IEEE Trans. Commun.,}  vol. 65, no. 10, pp. 4334 -- 4347, Oct. 2017.

\bibitem{vlc3}
L. Yin, W. O. Popoola, X. Wu, and H. Haas, "Performance evaluation of non-orthogonal multiple access in visible light communication,"  \emph{IEEE Trans. Commun.,} vol. 64, no. 12, pp. 5162--5175, Dec. 2016.

\bibitem{vlc4}
H. Marshoud, P. C. Sofotasios, S. Muhaidat, G. K. Karagiannidis, and B. S. Sharif, "On the performance of visible light communication systems with non-orthogonal multiple access," \emph{IEEE Trans. Wireless  Commun.,} vol. 16, no. 10, pp. 6350--6364, Oct. 2017.


\end{thebibliography}
\end{document}